# A model of weak viscoelastic nematodynamics

Arkady I. Leonov, Department of Polymer Engineering, The University of Akron,

Akron, OH 44325-0301, U.S.A. E-mail: leonov@uakron.edu

Abstract

The paper develops a continuum theory of weak viscoelastic nematodynamics of Maxwell type. It may describe the molecular elasticity effects in mono-domain flows of liquid crystalline polymers as well as the viscoelastic effects in suspensions of uniaxially symmetric particles in polymer fluids. Along with viscoelastic and nematic kinematics, the theory employs a general form of weakly elastic thermodynamic potential and the Leslie-Ericksen-Parodi type constitutive equations for viscous nematic liquids, while ignoring inertia effects and the Frank (orientation) elasticity in liquid crystal polymers. In general case, even the simplest Maxwell model has many basic parameters. Nevertheless, recently discovered algebraic properties of nematic operations reveal a general structure of the theory and present it in a simple form. It is shown that the evolution equation for director is also viscoelastic. An example of magnetization exemplifies the action of non-symmetric stresses. When the magnetic field is absent, the theory is simplified to the symmetric, fluid mechanical case with relaxation properties for both the stress and director. Our recent analyses of elastic and viscous soft deformation modes are also extended to the viscoelastic case. The occurrence of possible soft modes minimizes both the free energy and dissipation, and also significantly decreases the number of material parameters. In symmetric linear case, the theory is explicitly presented in terms of anisotropic linear memory functionals. Several analytical results demonstrate a rich behavior predicted by the developed model for steady and unsteady flows in simple shearing and simple elongation.

## I. Introduction

For various liquid crystalline (LC) materials, the internal rotational degree of freedom results in occurrence of internal couples and non-symmetry of stresses. Additionally, in case of LC polymers (LCP's) a partial flexibility of polymer chains, or molecular elasticity, can also be important. It means that along with nematic phenomena, typical for low molecular weight LC's, the viscoelasticity of LCP's could not be generally ignored. The viscoelastic characteristics of these systems, being anisotropic, are defined by essentially larger number of constitutive scalar parameters, as compared to their isotropic polymeric counterparts. Their experimental determination is a challenging problem.

The term *nematodynamics* introduced in the text by de Gennes and Prost (1974) means a specific set of problems for deformation and flow of the nematic systems under stress and external (magnetic and electrical) fields, which could be solved or analyzed using specific macroscopic field equations. The nematodynamic studies are important for processing of nematics and prediction of their post-processing properties. After more than twenty-five years of research, the polymer nematics still have a vast industrial potential. However, the complicated behavior of these systems, currently poorly understood, prevents the progress in processing of these systems and predicting the properties of post-processed products.

Two types of theories, continuum and molecular, attacked the problem of modeling LCP properties. While the continuum theories try to establish a general framework with minimum assumptions about the molecular structures of LCP's, the molecular approaches employ very specific assumptions of structure of these polymers, but unlike the continuum approaches, operate with few molecular parameters. These theories, however well separated, are not contradictory but supplemental.

Larson and Mead (1989) extended the Ericksen flow theory of LMV LC's to viscoelastic case, using *ad hoc* introduced linear viscoelastic operators instead of Ericksen's viscosities. Volkov and Kulichikhin (1990, 2000a,b) proposed a continuum, non-thermodynamic approach to weak anisotropic viscoelasticity of Maxwell type with internal rotations, based on symmetry arguments. Pleiner and Brand (1991, 1992) developed a thermodynamic theory of linear anisotropic viscoelasticity for LC polymers, involving gradients of state variables. Rey (1995a,b) also applied a thermodynamic approach for describing weak nonlinear phenomena in flow of LCP's. He obtained, however, doubtful results regarding the description of asymmetric stress.

A lot of theoretical effort was made to develop molecular theories that could model the lyotropic LC polymers. The theories by Marrucci and Greco (1993), Larson and co-authors (1998), and Feng et al (2000) typically use and elaborate the molecular long rigid rod approach, proposed by Doi (1981) and extended in the text by Doi and Edwards (1986). Also, B. Edwards *et al* (1990) applied the Poisson-Bracket continuum approach for developing constitutive equations (CE's) for LCP's [see also the text by Beris and Edwards (1999)]. This theory is reduced to the Doi theory in the homogeneous (mono-domain) limit. All the theories mentioned in this paragraph employ the same state variables as in case of LMW liquid crystals, i.e. the director $\underline{n}$ (or the second rank order tensor), and the director's space gradient $\underline{\nabla n}$. Additionally, one should mention the Rouse-like molecular theories of LCP's developed by Volkov and Kulichikhin (1994) and Long and Morse (2002). These theories take into account the partial flexibility of macromolecular chains in LCP and [in paper by Volkov and Kulichikhin (1994)] anisotropic macromolecular environment.

In order to theoretically describe the LCP properties, some additional specific problems have to be addressed. The first is a possible involvement of the Frank elasticity in polymer nematodynamics. In equilibrium case valid for liquid crystalline elastomers (LCE's), the Frank and molecular polymer elasticities have well separated space scales. Their crossover, a "characteristic scale" $l_*$ is evaluated as: $l_* = \sqrt{K/G}$ [Warner and Terentjev (2003)]. Employing here $G >\sim 10^6$ dyn/cm$^2$ as a typical rubber-like modulus and $K \sim 10^{-7}$ dyne as a typical value of the Frank moduli, gives: $l_* <\sim 10^{-6} cm = 10 nm$. It means that in the macroscopic scales the effects of Frank elasticity on the nematodynamics of elastomers should be ignored.

Although the above expression for $l_*$ could also be applied for evaluating the dominance of molecular (or "instant") elasticity in LCP's, non-equilibrium effects in these polymer systems do not allow ignoring the Frank elasticity. The most important is the existence of multi-domain "textures" for many LCP's at rest, which affects the slow (low Deborah number) flows of LCP's [e.g. see Larson (1998)]. The results of experiments by Odell *et al* (1993) show that the mono-domain description valid for flow of LCP's in relatively strong stress/external fields, acquire near the equilibrium some small-scale space periodicity due to the action of Frank elasticity or weak fluctuations, ignored in the mono-domain approaches. And *vice versa*, the polymer nematics usually forget their texture properties under action of relatively strong stress or external fields, which makes possible their mono-domain description. Complimentary to expression for the characteristic space scale $l_*$, there is the following scaling evaluation $t_\delta = \delta^2\eta / K$ for a characteristic time $t_\delta$ of spontaneous disorientation when forming a texture with characteristic domain size $\delta$. Here $\eta$ is a viscosity, of LCP near the equilibrium. Using the value $\eta \approx 10^4$ dn/cm$^2$ and $\delta \approx 10^{-4}$ cm for a typical multi-domain texture scale, leads to realistic evaluation, $t_\delta \approx 10^3$ sec. Odell at al (1993) found two time scales, one for fast stress relaxation and second for the slow disorientation during a period of ~ $10^3$ sec.

The possible effect of stress or external field on isotropic-nematic phase transition represent additional problem. In equilibrium, this phase transition can be commonly described by the well-known Landau phenomenology, or more specifically (however, less reliably because of large fluctuations known in nematics) by the Maier-Saupe mean field theory [see de Genes and Prost (1993), and also recent papers by Pickett and Schweizer (2000a,b)]. Although molecular Doi theory [Doi and Edwards (1986)] or Ericksen phenomenology (1991) included the possible dependence of scalar/tensor order parameter on stress/external field, this parameter was found being independent of stress when testing the LCE theory for nematic elastomers [Warner and Terentjev (2003)].

The present paper develops a continuum theory of non-symmetric weak viscoelastic nematodynamics of Maxwell type, where the viscous and elastic stresses coincide with the total stress. The assumption of small transient (elastic) strains and relative rotations, employed in the theory, seems to be appropriate for LCP's with rigid enough macromolecules, and for slow flows of viscoelastic suspensions with uniaxially shaped particles. This theory utilizes a specific viscoelastic and nematic kinematics and ignores inertia effects and in case of LCP's, the effects of Frank elasticity. The occurrence of non-symmetric stresses is demonstrated on the example of external magnetic field. It is shown the equation derived for the director evolution in weakly nonlinear case is also viscoelastic. The theory is simplified in "fluid mechanical" symmetric case. The analysis of possible soft nematic deformation modes for elastic and viscous nematics [Leonov and Volkov (2004a,b)] is extended in this paper to the viscoelastic case. In the symmetric linear case the theory is explicitly presented in the form of linear memory functionals. Several analytical solutions of nematodynamic equations, obtained for steady and unsteady simple shearing and simple elongation flows, demonstrate a variety of new possible effects. In applications to LCP, the above estimations show that the mono-domain approach developed in the present paper can describe the space variations of field variables in the scales exceeding one micron.

## II. Non-symmetricviscoelastic nematodynamics of Maxwell type

Bearing in mind that the main objective of this paper is developing viscoelastic, nematic CE's, we avoid discussing here the well-known general balance equations of momentum moment for nematics, as well as the related rotational inertia effects [e.g. see the texts by de Gennes and Prost (1974) and Kleman (1984), and in more detail, the paper by Leonov and Volkov (2002)].

2.1. Kinematical relations

We use in the following the viscoelastic kinematics based on the decomposition, $\underline{\underline{F}} = \underline{\underline{F}}_e \cdot \underline{\underline{F}}_p$, of full strain gradient $\underline{\underline{F}}$ into elastic (transient) $\underline{\underline{F}}_e$ and inelastic (viscous) $\underline{\underline{F}}_p$ parts [e.g. see Leonov (1987)]. Differentiating this equation with respect to time with further right multiplication by $\underline{\underline{F}}$, yields the Eulerian kinematical rate equation: $\underline{\underline{\nabla \mathrm{v}}} \equiv \dot{\underline{\underline{F}}} \cdot \underline{\underline{F}}^{-1} = \dot{\underline{\underline{F}}}_e \cdot \underline{\underline{F}}_e^{-1} + \underline{\underline{F}}_e \cdot \dot{\underline{\underline{F}}}_p \cdot \underline{\underline{F}}_p^{-1} \cdot \underline{\underline{F}}_e^{-1}$. Here $\underline{\underline{\nabla \mathrm{v}}}$ is the velocity gradient tensor. We now assume the elastic transient strain $\underline{\underline{\varepsilon}}$ and (body) elastic rotation $\underline{\underline{\Omega}}_e^b$ to be small, i.e. $\underline{\underline{F}}_e \approx \underline{\underline{\delta}} + \underline{\underline{\varepsilon}} + \underline{\underline{\Omega}}_e^b$, where $\underline{\underline{\delta}}$ is the unit tensor. Inserting these formulae into the above Eulerian kinematical relation results in the simplified rate equations:

$$\overset{0}{\underline{\underline{\varepsilon}}} + \underline{\underline{e}}_p = \underline{\underline{e}} \tag{1}$$

$$\overset{0}{\underline{\underline{\Omega}}}{}_e^b + \underline{\underline{\omega}}_p^b = \underline{\underline{\omega}}^b \tag{$2_1$}$$

Here the symbol $^0$ stands for Jaumann co-rotation time derivative, $\underline{\underline{e}} = 1/2[\underline{\underline{\nabla \mathrm{v}}} + (\underline{\underline{\nabla \mathrm{v}}})^T]$ and $\underline{\underline{\omega}}^b = 1/2[\underline{\underline{\nabla \mathrm{v}}} - (\underline{\underline{\nabla \mathrm{v}}})^T]$ are respectfully the full strain rate and vorticity of "body", so that $\underline{\underline{\nabla \mathrm{v}}} = \underline{\underline{e}} + \underline{\underline{\omega}}^b$, whereas $\underline{\underline{e}}_p$ and $\underline{\underline{\omega}}_p^b$ being respectfully the irreversible (viscous) strain rate and vorticity. Kinematical rate equations (1) and ($2_1$) could approximately describe only small elastic strain/vorticity superimposed on large inelastic strain/vorticity [e.g. see Gorodtsov and Leonov (1968)]. This however, might be a quite realistic feature for flows of weakly elastic nematic liquids. For simplicity we will consider below the incompressible situation when the tensors $\underline{\underline{\varepsilon}}$, $\underline{\underline{e}}_p$ and $\underline{\underline{e}}$ are traceless.

The kinematics of internal rotations starts from the equation, $\underline{n} = \underline{\underline{Q}} \cdot \underline{n}_0$, which relates the actual, $\underline{n}$ and initial $\underline{n}_0$ positions of director via orthogonal transformation with tensor $\underline{\underline{Q}}(\underline{x}, t)$. Differentiating this relation with respect to time yields: $\dot{\underline{n}} = \underline{\underline{\omega}}^i \cdot \underline{n}_0$, where $\underline{\underline{\omega}}^i = \dot{\underline{\underline{Q}}} \cdot \underline{\underline{Q}}^{-1}$ is the "vorticity" of internal rotations and overdot means the "material time derivative". Decomposing the orthogonal tensor $\underline{\underline{Q}}$ in elastic $\underline{\underline{Q}}_e$ and inelastic $\underline{\underline{Q}}_p$ orthogonal parts, results in the kinematical relation for internal rotations: $\underline{\underline{\omega}}^i = \dot{\underline{\underline{Q}}}_e \cdot \underline{\underline{Q}}_e^{-1} + \underline{\underline{Q}}_e \cdot \dot{\underline{\underline{Q}}}_p \cdot \underline{\underline{Q}}_p^{-1} \cdot \underline{\underline{Q}}_e^{-1} = \underline{\underline{\omega}}_e^i + \underline{\underline{\omega}}_p^i$. Using the expression, $\underline{\underline{Q}}_e = \exp \underline{\underline{\Omega}}_e^i$ where $\underline{\underline{\Omega}}_e^i$ is the asymmetric tensor of internal rotations assumed to be small, results in the additional rate equation describing internal rotations for nematic continua:

$$\overset{0}{\underline{\underline{\Omega}}}{}_e^i + \underline{\underline{\omega}}_p^i = \underline{\underline{\omega}}^i . \tag{$2_2$}$$

Extracting ($2_2$) from ($2_1$) yields the equation for *relative rotations* in weakly nonlinear nematic viscoelastic liquid:

$$\overset{0}{\underline{\underline{\Omega}}}_e + \underline{\underline{\omega}}_p = \underline{\underline{\omega}} \quad \left( \underline{\underline{\Omega}}_e = \underline{\underline{\Omega}}_e^b - \underline{\underline{\Omega}}_e^i,\ \underline{\underline{\omega}}_p = \underline{\underline{\omega}}_p^b - \underline{\underline{\omega}}_p^i,\ \underline{\underline{\omega}} = \underline{\underline{\omega}}^b - \underline{\underline{\omega}}^i \right). \qquad (2_3)$$

Additionally, there is the well-known kinematical relation, describing the rigid rotations of director $\underline{n}$ as:

$$\underline{\underline{\omega}} \cdot \underline{n} = \overset{0}{\underline{n}}. \qquad (3)$$

Equations (1)-(3) have been used in the linear limit by Pleiner and Brandt (1991,1992).

We finally introduce the kinematical variables, convenient for characterizing a combined effect of viscoelastic deformations and relative rotations:

$$\overset{0}{\underline{\underline{\Gamma}}}_e + \underline{\underline{\gamma}}_p = \underline{\underline{\gamma}} \quad \left( \overset{0}{\underline{\underline{\Gamma}}}_e = \overset{0}{\underline{\underline{\varepsilon}}} + \overset{0}{\underline{\underline{\Omega}}}_e,\ \underline{\underline{\gamma}}_p = \underline{\underline{e}}_p + \underline{\underline{\omega}}_p,\ \underline{\underline{\gamma}} = \underline{\underline{e}} + \underline{\underline{\omega}} \right). \qquad (4)$$

Note that kinematical tensors in (4) written as sums of symmetric and anti-symmetric components, should not be viewed as deformation gradients and velocity gradients, because their asymmetric parts describe the *relative* rotations.

It should be noted that in the equations (1), $(2_1)$ and $(2_2)$, the kinematical variables $\underline{\underline{\varepsilon}}$, $\underline{\underline{\Omega}}_e^b$ and $\underline{\underline{\Omega}}_e^i$ are not infinitesimal but finite-small. E.g. $\underline{\underline{\varepsilon}}$ is in fact the small Hencky elastic strain as well as the asymmetric tensors $\underline{\underline{\Omega}}_e^b$ and $\underline{\underline{\Omega}}_e^i$ are the logariphmic measures of corresponding asymmetric tensors characterizing small body and internal elastic rotations. Only in this case one can justify the occurrence of the Jaumann tensor time derivatives in the above kinematical relations.

2.2. Thermodynamics and constitutive relations

To describe the *quasi-equilibrium effects* in weak nematic viscoelasticity, valid in case of LCP, for the mono-domain situation, we will use as the most general, the following elastic potential (Helmholtz free energy density):

$$f = 1/2G_0 \left|\underline{\underline{\varepsilon}}\right|^2 + G_1 \underline{n}\underline{n} : \underline{\underline{\varepsilon}}^2 + G_2 (\underline{n}\underline{n} : \underline{\underline{\varepsilon}})^2 - 2G_3 \underline{n}\underline{n} : (\underline{\underline{\varepsilon}} \cdot \underline{\underline{\Omega}}_e) - G_5 \underline{n}\underline{n} : \underline{\underline{\Omega}}_e^2 \qquad (5)$$

Here $G_k$ are nematic moduli. Leonov and Volkov (2004a) employed the potential (5) for analyzing weak nematic elasticity, and discussed its relation to the de Gennes' (1980) potential. Note that the Cosserat, non-nematic isotropic term $\sim tr\underline{\underline{\Omega}}_e^2$ is omitted in (5) because its values in the isotropic case are typically very small.

Underlying the elasticity concept is the fundamental demand that free energy in equilibrium, i.e. in the elastic case, is at minimum. It should be noted that in the mono-domain case of nematic elasticity, the value of director is found either from the direct action of external fields in static cases and/or from the additional kinematical relation (3) for continuous elastic deformations with $\underline{\underline{\omega}}_p = 0$ [Leonov and Volkov (2004a)]. It means that in the mono-domain elastic case, director should not be considered as a state variable. The same happens with the description of slow flows of viscous nematics by the principle of minimum of dissipation, when the director field is not varied but is found from an additional kinematical relation [Leonov (2005)].

Among several sources of stress asymmetry, such as inertial effects of internal rotations, orientation (Frank) elasticity and the Cosserat/Born isotropic couples, only the most important action of external magnetic field $\underline{H}$ is taken into account in the mono-domain case considered below. Under the common assumptions, the magnetic field is potential, with the potential function $\Psi = -1/2\underline{\underline{\chi}}(\underline{n}):\underline{H}\underline{H}$. Here $\underline{\underline{\chi}}(\underline{n}) = \chi_{\perp}\underline{\underline{\delta}} + \chi_a \underline{n}\underline{n}$ is the susceptibility tensor, $\underline{\underline{\delta}}$ is the unit tensor, $\chi_{\|}$ and $\chi_{\|}$ are the susceptibilities parallel and perpendicular to the director, with $\chi_a = \chi_{\|} - \chi_{\perp}$ being the magnetic anisotropy. The body couple (or "effective magnetic field") is defined as:

$$\underline{h} = -\partial\Psi / \partial\underline{n} = \chi_a \underline{H}(\underline{n}\cdot\underline{H}) . \tag{6}$$

The equilibrium equation for internal torque in magnetic field is:

$$\underline{\underline{\sigma}}^a \equiv \underline{\underline{h}}^a = 1/2(\underline{h}\underline{n} - \underline{n}\underline{h}), \quad \underline{\underline{\sigma}}^a \cdot \underline{n} = 1/2[\underline{h} - \underline{n}(\underline{h}\cdot\underline{n})] . \tag{7}$$

Here $\underline{\underline{\sigma}}^a$ is asymmetric part of the stress tensor. The relation (7) shows that in the absence of magnetic field, the stress tensor in the present approach is symmetric.

We now use the well-known expressions for the *entropy production* $P_s$ in non-equilibrium systems [e.g. see de Groot and Mazur (1962)]:

$$TP_s = -\underline{q}\cdot\underline{\nabla}T + \underline{\underline{\sigma}}^s : \underline{\underline{e}} + \underline{\underline{\sigma}}^a : \underline{\underline{\omega}} - df\,\big|_T / dt. \tag{8$_1$}$$

Here $\underline{q}$ is the thermal flux, $T$ is the temperature, $\underline{\underline{\sigma}}^s$ and $\underline{\underline{\sigma}}^a$ are the symmetric and asymmetric parts of the extra stress tensor, respectively. Due to the second law of thermodynamics $P_s$ is strictly positive for non-equilibrium processes and vanishes in the equilibrium. Calculating the last term in ($8_1$) with the use of equations (1), ($2_3$), (3)-(6) yields:

$$TP_s = -\underline{q}\cdot\underline{\nabla}T + \underline{\underline{\sigma}}^s_p : \underline{\underline{e}} + \underline{\underline{\sigma}}^a_p : \underline{\underline{\omega}} + \underline{\underline{\sigma}}^s_e : \underline{\underline{e}}_p + \underline{\underline{\sigma}}^a_e : \underline{\underline{\omega}}_p , \tag{8$_2$}$$

Here,

$$\underline{\underline{\sigma}}^s_p \equiv \underline{\underline{\sigma}}^s - \underline{\underline{\sigma}}^s_e, \quad \underline{\underline{\sigma}}^a_p \equiv \underline{\underline{\sigma}}^a - \underline{\underline{\sigma}}^a_e, \quad \underline{\underline{\sigma}}^s_e = \partial f / \partial\underline{\underline{\varepsilon}}, \quad \underline{\underline{\sigma}}^a_e = \partial f / \partial\underline{\underline{\Omega}}_e . \tag{9}$$

In (9), $\underline{\underline{\sigma}}^s_e$ and $\underline{\underline{\sigma}}^a_e$ are equilibrium symmetric and asymmetric parts of the extra stress tensor, represented via elastic potential (5) as:

$$\begin{aligned}\underline{\underline{\sigma}}^s &= \partial f / \partial\underline{\underline{\varepsilon}} = G_0\underline{\underline{\varepsilon}} + G_1[\underline{n}\underline{n}\cdot\underline{\underline{\varepsilon}} + \underline{\underline{\varepsilon}}\cdot\underline{n}\underline{n} - 2nn(\underline{\underline{\varepsilon}}:\underline{n}\underline{n})] + 2(G_1+G_2)(\underline{n}\underline{n} - \underline{\underline{\delta}}/3)(\underline{\underline{\varepsilon}}:\underline{n}\underline{n}) \\ &\quad + G_3(\underline{n}\underline{n}\cdot\underline{\underline{\Omega}}_e - \underline{\underline{\Omega}}_e\cdot\underline{n}\underline{n})\end{aligned} \tag{10$_1$}$$

$$\underline{\underline{\sigma}}^a = \partial f / \partial\underline{\underline{\Omega}}_e = -G_4(\underline{n}\underline{n}\cdot\underline{\underline{\varepsilon}} - \underline{\underline{\varepsilon}}\cdot\underline{n}\underline{n}) + G_5(\underline{n}\underline{n}\cdot\underline{\underline{\Omega}}_e + \underline{\underline{\Omega}}_e\cdot\underline{n}\underline{n}). \tag{10$_2$}$$

Here $\underline{\underline{\sigma}}^a = \underline{\underline{h}}^a$ if the inertial effects of internal rotations are ignored.

Expression ($8_2$) for the entropy production demonstrates the various possible sources of non-equilibrium in nematic viscoelastic liquid: (i) non-isothermality, (ii) the dissipation produced by the non-equilibrium extra stress, $\underline{\underline{\sigma}}^s_p + \underline{\underline{\sigma}}^a_p$, on the total strain rate $\underline{\underline{e}}$ and relative vorticity $\underline{\underline{\omega}}$, and (iii) the dissipation produced by the equilibrium extra stress, $\underline{\underline{\sigma}}^s_e + \underline{\underline{\sigma}}^a_e$, on the irreversible strain rate $\underline{\underline{e}}_p$ and irreversible relative vorticity $\underline{\underline{\omega}}_p$.

The general quasi-linear constitutive relation between the thermodynamic forces $\{\underline{\nabla}T, \underline{\underline{\sigma}}^s_p, \underline{\underline{\sigma}}^a_p, \underline{\underline{\sigma}}^s_e, \underline{\underline{\sigma}}^a_e\}$ and fluxes $\{\underline{q}, \underline{\underline{e}}, \underline{\underline{\omega}}, \underline{\underline{e}}_p, \underline{\underline{\omega}}_p\}$ can be readily established with account for the Onsager symmetry, via the kinetic matrices $\mathbf{A}^{mn}$ representing some tensors depending on director $\underline{n}$. However, this general relation seems to be intractable. Therefore we analyze in the following only the simplest, Maxwell type, CE for viscoelastic nematic liquids, when in ($8_2$) $\underline{\underline{\sigma}}_p^s = 0$, $\underline{\underline{\sigma}}_p^a = 0$, i.e. $\underline{\underline{\sigma}}_e^s = \underline{\underline{\sigma}}^s$ and $\underline{\underline{\sigma}}_e^a = \underline{\underline{\sigma}}^a$. Then the relation ($8_2$) for entropy production takes the form:

$$TP_s = -\underline{q}\cdot\underline{\nabla}T + \underline{\underline{\sigma}}^s : \underline{\underline{e}}_p + \underline{\underline{\sigma}}^a : \underline{\underline{\omega}}_p. \qquad (11)$$

Due to (11), the constitutive relations between the irreversible kinematical variables and stress are of the LEP type:

$$\begin{aligned}\underline{\underline{\sigma}}^s = \eta_0\underline{\underline{e}}_p + \eta_1[\underline{nn}\cdot\underline{\underline{e}}_p + \underline{\underline{e}}_p\cdot\underline{nn} - 2\underline{nn}(\underline{nn}:\underline{\underline{e}}_p)] + 2(\eta_1+\eta_2)(\underline{nn}-\underline{\underline{\delta}}/3)(\underline{nn}:\underline{\underline{e}}_p) \\ + \eta_3(\underline{nn}\cdot\underline{\underline{\omega}}_p - \underline{\underline{\omega}}_p\cdot\underline{nn})\end{aligned}, \qquad (12_1)$$

$$\underline{\underline{\sigma}}^a = -\eta_4(\underline{nn}\cdot\underline{\underline{e}}_p - \underline{\underline{e}}_p\cdot\underline{nn}) + \eta_5(\underline{nn}\cdot\underline{\underline{\omega}}_p + \underline{\underline{\omega}}_p\cdot\underline{nn}) \quad (\eta_4 = -\eta_3). \qquad (12_2)$$

As in recent paper by Leonov and Volkov (2004,b), the coupled "viscous" CE's ($12_{1,2}$) are written in separate symmetric and asymmetric forms. The Onsager relation, $\eta_4 = -\eta_3$, between the coupled terms in equations ($12_1$) and ($12_2$) has also been used here. Similarly to the elastic part, the isotropic, non-nematic Borns term $\sim \underline{\underline{\omega}}_p$ is omitted in ($12_2$) because its values in the isotropic case are typically very small. The relations between the Leslie-Ericksen parameters $\alpha_k$ and the "viscosities" $\eta_k$ in ($12_{1,2}$) are:

$$\alpha_1 = 2\eta_2,\ \alpha_2 = -\eta_3 - \eta_4,\ \alpha_3 = -\eta_3 + \eta_4,\ \alpha_4 = \eta_0,\ \alpha_5 = \eta_1 + \eta_3,\ \alpha_6 = \eta_1 - \eta_3.$$

Here additional Parodi equality, $\alpha_2 + \alpha_3 = \alpha_6 - \alpha_5 \ \ (= -2\eta_3)$, is identically satisfied.

Unlike the nematic ferrofluids, in case of LC's and LCP's, the dependence of kinetic coefficients on magnetic field is commonly ignored [e.g. see Lubensky (1973) and Jarkova *et al* (2001)]. Therefore viscosities $\eta_k$ are considered as $\underline{H}$ independent.

Finally, the anisotropic Fourier law has the form:

$$\underline{q} = -\underline{\underline{\kappa}}(\underline{n},T)\cdot\underline{\nabla}T\ ; \quad \underline{\underline{\kappa}}(\underline{n},T) = \kappa_\perp(T)\underline{\underline{\delta}} + \kappa_a(T)\underline{nn} \quad (\kappa_a = \kappa_\parallel - \kappa_\perp). \qquad (13)$$

Demanding the thermodynamic stability for the elastic potential (5) results in the necessary and sufficient stability conditions [Leonov and Volkov ( 2004a)]:

$$G_0 > 0;\ \ G_5 > 0;\ \ G_0 + G_1 > 0;\ \ 3/4G_0 + G_1 + G_2 > 0;\ \ (G_0 + G_1)G_5 > G_3^2. \qquad (14_1)$$

The thermodynamic stability conditions for dissipation in (8) in incompressible case, established by Leonov and Volkov (2004b), are the same as in ($14_1$) with substitution $G_k \to \eta_k$, i.e.

$$\eta_0 > 0;\ \ \eta_5 > 0;\ \ \eta_0 + \eta_1 > 0;\ 3/4\eta_0 + \eta_1 + \eta_2 > 0;\ \ (\eta_0 + \eta_1)\eta_5 > \eta_3^2. \qquad (14_2)$$

Additionally, to avoid the degeneration of CE's it is further assumed that $G_k \neq 0$ and $\eta_k \neq 0$.

The thermodynamic stability conditions for the thermal processes are:

$$\kappa_\perp > 0, \quad \kappa_\parallel > 0\ \ (\kappa_\perp + \kappa_a > 0). \qquad (15)$$

Substituting now CE's ($14_{1,2}$) and (15) into the expression for entropy production (11) reduces the latter to the quadratic form:

$$TP_s = \underline{\underline{\kappa}}(\underline{n},T)\cdot\underline{\nabla}T\underline{\nabla}T + \eta_0\left|\underline{\underline{e}}_p\right|^2 + 2\eta_1\underline{nn}:\underline{\underline{e}}_p^2 + 2\eta_2(\underline{nn}:\underline{\underline{e}}_p)^2 - 4\eta_3\underline{nn}:(\underline{\underline{e}}_p\cdot\underline{\underline{\omega}}_p) - 2\eta_5\underline{nn}:\underline{\underline{\omega}}_p^2. \qquad (16)$$

Due to the stability constraints ($14_{1,2}$) and (15) the quadratic form in (16) is positively definite.

### 2.3. Evolution equations for elastic (transient) strain $\underline{\underline{\varepsilon}}$ and rotation $\underline{\underline{\Omega}}_e$

The tensors $\underline{\underline{\omega}}_p$ and $\underline{\underline{e}}_p$ in (2)-(4), characterizing irreversible kinematical effects, are considered below as the rate measures of deviation of "quasi-equilibrium" state variables $\underline{\underline{\varepsilon}}$ and $\underline{\underline{\Omega}}_e$ from their full kinematical parts. Excluding $\underline{\underline{\omega}}_p$ and $\underline{\underline{e}}_p$ from the above constitutive relation yields a coupled set of equations for evolution of the hidden thermodynamic parameters $\underline{\underline{\Omega}}_e$ and $\underline{\underline{\varepsilon}}$. In order to do that the irreversible kinematical tensor $\underline{\underline{\gamma}}_p = \underline{\underline{e}}_p + \underline{\underline{\omega}}_p$ has to be expressed in the kinematical equation (4) via the quasi-equilibrium tensor parameter $\underline{\underline{\Gamma}}_e = \underline{\underline{\varepsilon}} + \underline{\underline{\Omega}}_e$. Since in the Maxwell liquid the stress is the same in viscous and elastic modes, equalizing the symmetric stress in equations ($10_1$) and ($12_1$) as well as asymmetric stress in equations ($10_2$) and ($12_2$) yields two equations, which can be written in the operator form as: $\underline{\underline{\sigma}} = \mathbf{G}(\underline{n})\bullet\underline{\underline{\Gamma}}_e = \boldsymbol{\eta}(\underline{n})\bullet\underline{\underline{\gamma}}_p$. Here $\mathbf{G}(\underline{n})$ and $\boldsymbol{\eta}(\underline{n})$ are some operators of moduli and viscosity, respectively, represented by their 4th rank tensors. The algebraic properties of these operators have been recently established [Leonov (2004)]. It was shown that under the stability conditions ($14_{1,2}$) the inverse operators for $\mathbf{G}(\underline{n})$ and $\boldsymbol{\eta}(\underline{n})$ do exist and have explicit expressions. Then one can express $\underline{\underline{\gamma}}_p$ via $\underline{\underline{\Gamma}}_e$ and *vice versa* as: $\underline{\underline{\gamma}}_p = \mathbf{s}(\underline{n})\bullet\underline{\underline{\Gamma}}_e$ and $\underline{\underline{\Gamma}}_e = \boldsymbol{\theta}(\underline{n})\bullet\underline{\underline{\gamma}}_p$. Here $\mathbf{s}(\underline{n}) = \boldsymbol{\eta}^{-1}(\underline{n})\bullet\mathbf{G}(\underline{n})$ ) and $\boldsymbol{\theta}(\underline{n}) = \mathbf{G}^{-1}(\underline{n})\bullet\boldsymbol{\eta}(\underline{n})$ are the fourth rank tensors of relaxation frequency and relaxation, respectively. Substituting this result in (4) yields the operator form of evolution equation: $\dot{\underline{\underline{\Gamma}}}_e + \mathbf{s}(\underline{n})\bullet\underline{\underline{\Gamma}}_e = \underline{\underline{\gamma}}$, which is presented as:

$$\overset{0}{\underline{\underline{\varepsilon}}} + s_0\underline{\underline{\varepsilon}} + s_1[\underline{nn}\cdot\underline{\underline{\varepsilon}} + \underline{\underline{\varepsilon}}\cdot\underline{nn} - 2nn(\underline{\underline{\varepsilon}}:\underline{nn})] + s_2(\underline{nn} - \underline{\underline{\delta}}/3)(\underline{\underline{\varepsilon}}:\underline{nn}) + s_3(\underline{nn}\cdot\underline{\underline{\Omega}}_e - \underline{\underline{\Omega}}_e\cdot\underline{nn}) = \underline{\underline{e}}$$

$$\overset{0}{\underline{\underline{\Omega}}}_e + s_4(\underline{nn}\cdot\underline{\underline{\varepsilon}} - \underline{\underline{\varepsilon}}\cdot\underline{nn}) + s_5(\underline{nn}\cdot\underline{\underline{\Omega}}_e + \underline{\underline{\Omega}}_e\cdot\underline{nn}) = \underline{\underline{\omega}} \qquad (17_{1,2})$$

Here the parameters $s_k$ are the basis parameters of relaxation frequency tensor $\mathbf{s}(\underline{n})$, which are presented via the basic parameters $G_k$ and $\eta_k$ in the model as:

$$s_0 = \frac{G_0}{\eta_0},\ \ s_1 = \frac{\eta_5(G_1\eta_0 - G_0\eta_1) + \eta_3(G_0\eta_3 - G_3\eta_0)}{\eta_0[\eta_5(\eta_0+\eta_1) - \eta_3^2]},\ \ s_2 = \frac{3}{2}\cdot\frac{(G_1+G_2)\eta_0 - G_0(\eta_1+\eta_2)}{\eta_0(3/4\eta_0 + \eta_1 + \eta_2)}$$
$$s_3 = \frac{G_3\eta_5 - G_5\eta_3}{\eta_5(\eta_0+\eta_1) - \eta_3^2},\ \ s_4 = \frac{G_3(\eta_0+\eta_1) - \eta_3(G_0+G_1)}{\eta_5(\eta_0+\eta_1) - \eta_3^2},\ \ s_5 = \frac{G_5(\eta_0+\eta_1) - G_3\eta_5}{\eta_5(\eta_0+\eta_1) - \eta_3^2} \qquad (18)$$

The basic parameters $\theta_k$ of relaxation tensor $\boldsymbol{\theta}(\underline{n})$ could be obtained from (18) using the double transformation: $\eta_k \to G_k,\ \ G_k \to \eta_k$.

Unlike the 4th rank tensors $\mathbf{G}(\underline{n})$ and $\boldsymbol{\eta}(\underline{n})$, tensor $\mathbf{s}(\underline{n})$ does not possess the Onsager symmetry, since generally $s_3 \neq s_4$. Equations ($12_{1,2}$) and ($17_{1,2}$) along with (18) form the closed set of CE's if parameter $\underline{n}$ is known. It should also be noted that the evolution equations ($17_{1,2}$) hold for non-isothermal flows.

### 2.4. Equations for elastic strain, director, and the stress tensor

We now introduce the normalized molecular field $\underline{h}_{+} = \underline{h}/G_5$ with its projection $\underline{h}_{+}^{\perp}$ on the normal to director, and the traceless normalized symmetric $\underline{\underline{h}}_{+}^{s}$ and asymmetric $\underline{\underline{h}}_{+}^{a}$ tensors. These quantities are defined using $\underline{h}_{+}$ as:

$$\underline{h}_{+}^{\perp} = \underline{h}_{+} - \underline{n}(\underline{h}_{+} \cdot \underline{n}), \quad \underline{\underline{h}}_{+}^{s} = 1/2[\underline{h}_{+}\underline{n} + \underline{n}\underline{h}_{+} - 2\underline{n}\underline{n}(\underline{h}_{+} \cdot \underline{n})], \quad \underline{\underline{h}}_{+}^{a} = 1/2(\underline{h}_{+}\underline{n} - \underline{n}\underline{h}_{+}). \qquad (19)$$

Equation ($10_2$) yields:

$$\underline{\underline{\Omega}}_{e} = \underline{\underline{h}}_{+}^{a} + \lambda_e(\underline{\underline{\varepsilon}} \cdot \underline{n}\underline{n} - \underline{n}\underline{n} \cdot \underline{\underline{\varepsilon}}) \qquad (\lambda_e = G_3/G_5). \qquad (20)$$

Equation ($12_2$) also yields the kinematical expression for $\underline{\underline{\omega}}_{p}$, which is similar to (20) with substitutions: $\underline{\underline{\Omega}}_{e} \to \underline{\underline{\omega}}_{p}$, $\underline{\underline{\varepsilon}} \to \underline{\underline{e}}_{p}$, $\lambda_e \to \lambda_{\text{v}} = \eta_3/\eta_5$. Here $\lambda_e$ and $\lambda_{\text{v}}$ are the elastic and viscous "tumbling" parameters.

Substituting (20) into ($17_1$) yields:

$$\overset{0}{\underline{\underline{\varepsilon}}} + s_0\underline{\underline{\varepsilon}} + \tilde{s}_1[\underline{n}\underline{n} \cdot \underline{\underline{\varepsilon}} + \underline{\underline{\varepsilon}} \cdot \underline{n}\underline{n} - 2nn(\underline{\underline{\varepsilon}} : \underline{n}\underline{n})] + s_2(\underline{n}\underline{n} - \underline{\underline{\delta}}/3)(\underline{\underline{\varepsilon}} : \underline{n}\underline{n}) = \underline{\underline{e}} + s_3\underline{\underline{h}}_{+}^{s} \qquad (21)$$
$$(\tilde{s}_1 = s_1 - \lambda_e s_3)$$

Note that equations (20) and (21) are also valid for non-isothermal flows.

Consider now isothermal situation. Substituting (20) into ($17_2$), right multiplied by $\underline{n}$, and taking into account ($2_3$) and (3) yields:

$$\overset{0}{\underline{\underline{\Omega}}}_{e} \cdot \underline{n} - (s_4 - \lambda_1 s_5)[\underline{\underline{\varepsilon}} \cdot \underline{n} - \underline{n}(\underline{\underline{\varepsilon}} : \underline{n}\underline{n})] = \overset{0}{\underline{n}} - \underline{h}_{+}^{\perp} s_5 .$$

Applying now Jaumann time derivative to equation (20) and substituting the result into the above equation with account of (21) yields:

$$\overset{0}{\underline{n}} - \lambda_e[\underline{\underline{\varepsilon}} \cdot \overset{0}{\underline{n}} - \overset{0}{\underline{n}}(\underline{\underline{\varepsilon}} : \underline{n}\underline{n}) - \underline{n}(\overset{0}{\underline{\underline{\varepsilon}}} : \underline{n}\underline{n})] = \mathbf{b}(\underline{n})\bullet(\lambda_e\underline{\underline{e}} + \xi_1\underline{\underline{\varepsilon}}) + \overset{0}{\underline{\underline{h}}_{+}^{a}} \cdot \underline{n} + \xi_2\underline{h}_{+}^{\perp} . \qquad (22_1)$$

Here the vector $\mathbf{b}(\underline{n})\bullet\underline{\underline{e}}$ is defined below in (23). In the weak nematic viscoelasticity where $|\underline{\underline{\varepsilon}}| << 1$ the second term in the right-hand side of ($22_1$) can be neglected as compared to the first one, if $\lambda_e \le O(1)$. Then equation ($22_1$) takes the simplified form:

$$\overset{0}{\underline{n}} \approx \mathbf{b}(\underline{n})\bullet(\lambda_e\underline{\underline{e}} + \xi_1\underline{\underline{\varepsilon}}) + \overset{0}{\underline{\underline{h}}_{+}^{a}} \cdot \underline{n} + \xi_2\underline{h}_{+}^{\perp} . \qquad (22_2)$$

In the equations ($22_{1,2}$), the vector $\mathbf{b}(\underline{n})\bullet\underline{\underline{e}}$ is defined as:

$$\left(\mathbf{b}(\underline{n})\bullet\underline{\underline{e}}\right)_i = b_{ijk}(\underline{n})e_{kj} = (\delta_{ij}n_k - n_i n_j n_k)e_{kj}, \quad \xi_1 = \lambda_e(s_0 + \hat{s}_1 + s_5) - s_4, \quad \xi_2 = s_5 - \lambda_e s_3 . \qquad (23)$$

Because of the presence of magnetic terms it is impossible to exclude the elastic strain $\underline{\underline{\varepsilon}}$ from ($22_2$) with the aid of (21), using only the inequality $|\underline{\underline{\varepsilon}}| << 1$. Therefore when magnetic field is imposed, equations (21) and (22) present *irreducible* coupled evolution equations for transient elastic strain $\underline{\underline{\varepsilon}}$ and director $\underline{n}$.

Finally, the equation for the total traceless extra stress tensor, $\underline{\underline{\sigma}} = \underline{\underline{\sigma}}^s + \underline{\underline{\sigma}}^a$, is readily obtained when substituting (20) in ($10_1$):

$$\underline{\underline{\sigma}} = G_0 \underline{\underline{\varepsilon}} + \hat{G}_1[\underline{nn}\cdot\underline{\underline{\varepsilon}} + \underline{\underline{\varepsilon}}\cdot\underline{nn} - 2nn(\underline{\underline{\varepsilon}}:\underline{nn})] + 2(G_1+G_2)(\underline{nn} - \underline{\underline{\delta}}/3)(\underline{\underline{\varepsilon}}:\underline{nn}) - G_3 \underline{\underline{h}}_+^s + G_5 \underline{\underline{h}}_+^a \quad (24)$$

$$(\hat{G}_1 = G_1 - G_3^2/G_5)$$

Equations (21), (22) and (24) are the closed set of Maxwell viscoelastic nematodynamics in the non-symmetric case when the magnetic field is presented.

In case of LCP's, the coupled evolution equations (21) and (22) describe the complicated process of orienting macromolecules caused by the two effects: (i) the local orientation of rigid segments in macromolecules and their conformational orientation-in-large due to the action of hydrodynamic field, and (ii) the local orientation of rigid segments in macromolecules due to the direct action of magnetic field. In two particular cases considered below the evolution equations (21) and (22) are simplified.

## III. Symmetric viscoelastic nematodynamics of Maxwell type

### 3.1. Symmetric nematic operators

This Section analyzes the cases when non-symmetric effects in viscoelastic nematodynamics are negligible. When the magnetic field is absent, the weakly nonlinear, case is analyzed in the Subsection 3.1. Another one, the Miescowicz viscoelasticity, valid when the magnetic field is strong enough, is linear and considered in the Subsection 3.3.

In order to simplify notations and make easier manipulations with awkward formulae, we now introduce some linear symmetric nematic operators whose theory was recently developed by the author [Leonov (2004)]. These operators are represented via symmetric fourth-rank, *transversely isotropic*, traceless tensors $\mathbf{A}(\underline{n}; a_0, a_1, a_2)$, whose action on various physical variables described by traceless symmetric second rank tensors $\underline{\underline{x}}$ is defined as:

$$\mathbf{A}(\underline{n}; a_0, a_1, a_2)\bullet\underline{\underline{x}} = a_0\underline{\underline{x}} + a_1[\underline{nn}\cdot\underline{\underline{x}} + \underline{\underline{x}}\cdot\underline{nn} - 2nn(\underline{\underline{x}}:\underline{nn})] + a_2(\underline{nn} - \underline{\underline{\delta}}/3)(\underline{\underline{x}}:\underline{nn}). \quad (25)$$

We will use below the following basic nematic operators:

(i) Modulus, $\hat{\mathbf{G}}(\underline{n}) = \mathbf{A}(\underline{n}; G_0, \hat{G}_1, 2\hat{G}_1 + 2\hat{G}_2)$; ($25_1$)

(ii) Viscosity, $\hat{\boldsymbol{\eta}}(\underline{n}) = \mathbf{A}(\underline{n}; \eta_0, \hat{\eta}_1, 2\hat{\eta}_1 + 2\hat{\eta}_2)$; ($25_2$)

(iii) Relaxation frequency, $\hat{\mathbf{s}}(\underline{n}) = \mathbf{A}(\underline{n}; s_0, \hat{s}_1, \hat{s}_2)$; and ($25_3$)

(iv) Relaxation time, $\hat{\boldsymbol{\theta}}(\underline{n}) = \mathbf{A}(\underline{n}; \theta_0, \hat{\theta}_1, \hat{\theta}_2)$. ($25_4$)

In formulae ($25_{1\text{-}4}$) we introduced, following Leonov and Volkov (2004a,b), the "reduced" moduli $\hat{G}_k$ and viscosities $\hat{\eta}_k$ as:

$$\hat{G}_1 = G_1 - G_3^2/G_5,\ \hat{G}_2 = G_2 + G_3^2/G_5;\ \ \hat{\eta}_1 = \eta_1 - \eta_3^2/\eta_5,\ \hat{\eta}_2 = \eta_2 + \eta_3^2/\eta_5. \quad (26_1)$$

Using ($26_1$) we also represent the relaxation frequencies $\hat{s}_k$ and relaxation times $\hat{\theta}_k$ in ($25_{3,4}$) as follows:

$$s_0 = \frac{G_0}{\eta_0},\quad \hat{s}_1 = \frac{\hat{G}_1\eta_0 - \hat{\eta}_1 G_0}{\eta_0(\eta_0 + \hat{\eta}_1)},\qquad \hat{s}_2 = \frac{3}{2}\cdot\frac{\eta_0(\hat{G}_1 + \hat{G}_2) - G_0(\hat{\eta}_1 + \hat{\eta}_2)}{\eta_0(3/4\eta_0 + \hat{\eta}_1 + \hat{\eta}_2)} \quad (26_2)$$

$$\theta_0 = \frac{\eta_0}{G_0}, \quad \hat{\theta}_1 = \frac{G_0\hat{\eta}_1 - \eta_0\hat{G}_1}{G_0(G_0+\hat{G}_1)}, \quad \hat{\theta}_2 = \theta_2 = \frac{3}{2}\cdot\frac{G_0(\hat{\eta}_1+\hat{\eta}_2)-\eta_0(\hat{G}_1+\hat{G}_2)}{G_0(3/4G_0+\hat{G}_1+\hat{G}_2)} \qquad (26_3)$$

It is seen that the relaxation times $\hat{\theta}_k$ in ($26_3$) are obtained from the relaxation frequencies $\hat{s}_k$ in ($26_2$) using the double transformations: $\hat{\eta}_k \to \hat{G}_k$, $\hat{G}_k \to \hat{\eta}_k$.

The (necessary and sufficient) conditions of *thermodynamic stability* ($15_{1,2}$) represented in terms of parameters $\hat{G}_k$ and $\hat{\eta}_k$ have the similar forms:

$$G_0 > 0,\ G_0 + \hat{G}_1 > 0,\ 3/4G_0 + \hat{G}_1 + \hat{G}_2 > 0 \qquad (27_1)$$

$$\eta_0 > 0,\ \eta_0 + \hat{\eta}_1 > 0,\ 3/4\eta_0 + \hat{\eta}_1 + \hat{\eta}_2 > 0. \qquad (27_2)$$

All the operators in (25) are commutative and due to the thermodynamic stability constraints ($27_{1,2}$) positively definite and have inverse. These operators have the same properties as in the isotropic case:

$$\hat{\boldsymbol{\theta}}(\underline{n}) = \hat{\mathbf{s}}^{-1}(\underline{n}) = \hat{\boldsymbol{\eta}}(\underline{n})\bullet\hat{\mathbf{G}}^{-1}(\underline{n}). \qquad (28)$$

3.2. Weak nematic viscoelasticity in the absence of magnetic field ($\underline{h} = 0$)

In this case, the stress-elastic strain relation (24) and the evolution equation (21) for transient elastic strain take the forms:

$$\underline{\underline{\sigma}} = \hat{\mathbf{G}}(\underline{n})\bullet\underline{\underline{\varepsilon}};\quad \overset{0}{\underline{\underline{\varepsilon}}} + \hat{\mathbf{s}}(\underline{n})\bullet\underline{\underline{\varepsilon}} = \underline{\underline{e}} \ \text{ or } \ \hat{\boldsymbol{\theta}}(\underline{n})\bullet\overset{0}{\underline{\underline{\varepsilon}}} + \underline{\underline{\varepsilon}} = \hat{\boldsymbol{\theta}}(\underline{n})\bullet\underline{\underline{e}} \qquad (29_{1,2,3})$$

Consider now the approximated evolution equation for director ($22_2$) with $\underline{h} = 0$. In this case the differentiation with frozen value of $\underline{n}$ is within the approximation of weak nematic viscoelasticity where $|\underline{\underline{\varepsilon}}| << 1$. When $\underline{h} = 0$, applying to ($22_2$) the operator defined in the left-hand side of ($29_2$), results in the evolution equation for director:

$$\theta^*\left(\overset{00}{\underline{n}} + \underline{n}\cdot\left|\overset{0}{\underline{n}}\right|^2\right) + \overset{0}{\underline{n}} = \mathbf{b}(\underline{n})\bullet(\lambda_e\theta^*\overset{0}{\underline{\underline{e}}} + \lambda_{\mathrm{v}}\underline{\underline{e}})$$
$$\theta^* = \frac{1}{\hat{s}_1+\hat{s}_2} = \frac{\eta_0+\hat{\eta}_1}{G_0+\hat{G}_1},\ \lambda_e = \frac{G_3}{G_5},\ \lambda_{\mathrm{v}} = \frac{\eta_3}{\eta_5}. \qquad (30)$$

Here $\lambda_e$ and $\lambda_{\mathrm{v}}$ are the elastic and viscous "tumbling" parameters. It is seen that the both sides of (30) vanish after scalar multiplication of this equation by $\underline{n}$. Relaxation character of this equation is caused by the "bulk" relaxation due to the viscoelastic nematic character of evolution equation in (29) for the transient elastic strain. Note that in particular cases when $\lambda_e = \lambda_{\mathrm{v}} \equiv \lambda$, as well as when either $\theta^* \to 0$, or $\theta^* \to \infty$, the evolution equation (30) reduces to the Ericksen equation, $\overset{0}{\underline{n}} = \lambda\mathbf{b}(\underline{n})\bullet\underline{\underline{e}}$.

Using the differentiation with frozen value of $\underline{n}$ one can also exclude the hidden parameter $\underline{\underline{\varepsilon}}$ from (29) to obtain the approximate evolution equation for symmetric extra stress, written in identical forms:

$$\overset{0}{\underline{\underline{\sigma}}} + \hat{\mathbf{s}}(\underline{n})\bullet\underline{\underline{\sigma}} = \hat{\mathbf{G}}(\underline{n}):\underline{\underline{e}} \ \text{ or } \ \hat{\boldsymbol{\theta}}(\underline{n})\bullet\overset{0}{\underline{\underline{\sigma}}} + \underline{\underline{\sigma}} = \hat{\boldsymbol{\eta}}(\underline{n})\bullet\underline{\underline{e}} \qquad (31_{1,2})$$

Equations (29) and (30), or in simplified case, (30) and (31), represent the closed set of symmetric weak viscoelastic nematodynamics of Maxwell type. Note that in the symmetric case, the CE ($31_2$) is the same as proposed by Volkov and Kulichikhin (1990, 2000a,b).

CE's (29) can also be directly obtained from the general formulation ($10_{1,2}$), ($12_{1,2}$) using two kinematical relations held in the absence of magnetic field ($\underline{\underline{\sigma}}^a = 0$):

$$\underline{\underline{\Omega}}_e = \lambda_e(\underline{\underline{\varepsilon}}\cdot\underline{n}\underline{n} - \underline{n}\underline{n}\cdot\underline{\underline{\varepsilon}}) \quad (\lambda_e = G_3/G_5),\ \underline{\underline{\omega}}_p = \lambda_v(\underline{\underline{e}}_p\cdot\underline{n}\underline{n} - \underline{n}\underline{n}\cdot\underline{\underline{e}}_p) \quad (\lambda_v = \eta_3/\eta_5) \quad (32_{1,2})$$

Substituting these relations into ($10_1$) and ($12_1$) yields the reduced formulation for the extra stress: $\underline{\underline{\sigma}} = \hat{\mathbf{G}}(\underline{n})\bullet\underline{\underline{\varepsilon}} = \hat{\boldsymbol{\eta}}(\underline{n})\bullet\underline{\underline{e}}_p$. Expressing from this dual equation $\underline{\underline{e}}_p$ via $\underline{\underline{\varepsilon}}$ and substituting the result in (4) yields equations (29).

3.2. Soft deformation modes in the weak Maxwell type nematodynamics

Leonov and Volkov (2004a,b) recently analyzed the possible soft deformation modes in elastic and viscous nematics, using the marginal stability approach. It was shown that the soft modes could exist only in the absence of external fields. This Section extends the above approaches to the viscoelastic nematic symmetric case under study. To simplify the notations, *we omit hereafter the overcaps in notations of* $\hat{G}_k$ and $\hat{\eta}_k$.

We consider below the general symmetric anisotropic case when the non-degenerating conditions $G_k \neq 0,\ \ \eta_k \neq 0\ \ (k = 0,1,2)$ are valid. The soft cases occur when the values of material parameters belong to the *marginal stability* boundaries in ($27_{1,2}$). Because of assumed non-degeneration of constitutive parameters, there are only four independent *marginal stability* conditions:

$$G_0 + G_1 = 0,\ \ \eta_0 + \eta_1 = 0;\ 3/4G_0 + G_1 + G_2 = 0;\ 3/4\eta_0 + \eta_1 + \eta_2 = 0. \qquad (33)$$

The first two equalities in (34) are related to the shearing soft modes, and another two to the extensional ones [Leonov and Volkov (2004a,b)]. We call *completely soft* the possible case when all four soft conditions in (35) are satisfied.

The *nearly marginal* or *semi-soft*, still stable situations happen when instead of (33) the four independent conditions are satisfied:

$$G_0 + G_1 = \delta_G G_0,\quad \eta_0 + \eta_1 = \delta_\eta \eta_0 \qquad (0 << \delta_G, \delta_\eta << 1) \qquad (34_1)$$

$$3/4G_0 + G_1 + G_2 = 3/2\kappa_G G_0,\quad 3/4\eta_0 + \eta_1 + \eta_2 = 3/2\kappa_\eta \eta_0 \quad (0 << \kappa_G, \kappa_\eta << 1) \quad (34_2)$$

Below we briefly follow the recent formal analyses of soft and semi-soft viscoelastic modes [Leonov (2004)], considering only the most important particular case when both the elastic and viscous behaviors in our nematic model are *completely* soft, i.e. when all the equalities (33) hold. In this case the expressions of extra stress tensor via elastic strain $\underline{\underline{\varepsilon}}$ or via viscous strain rate $\underline{\underline{e}}_p$ have the similar, one-parametric forms:

$$\underline{\underline{\sigma}}/G_0 = \boldsymbol{\alpha}(\underline{n})\bullet\underline{\underline{\varepsilon}},\quad \underline{\underline{\sigma}}/\eta_0 = \boldsymbol{\alpha}(\underline{n})\bullet\underline{\underline{e}}_p,\quad \boldsymbol{\alpha}(\underline{n}) = \mathbf{A}(\underline{n};1,-1,-3/2). \qquad (35_{1.2,3})$$

Here $\boldsymbol{\alpha}(\underline{n})$ is a numerical operator, defined in (25). Unlike the regular case the completely soft case is singular, meaning that generally the inverse operations in ($35_{1,2}$), expressing $\underline{\underline{\varepsilon}}$ or $\underline{\underline{e}}_p$ via $\underline{\underline{\sigma}}$, do not exist.

The free energy and dissipation in completely soft case take the similar form:

$$f/G_0 = 1/2\left|\underline{\underline{\varepsilon}}\right|^2 - \underline{n}\underline{n}:\underline{\underline{\varepsilon}}^2 + 1/4(\underline{n}\underline{n}:\underline{\underline{\varepsilon}})^2; \quad 2D/\eta_0 = 1/2\left|\underline{\underline{e}}_p\right|^2 - \underline{n}\underline{n}:\underline{\underline{e}}_p^2 + 1/4(\underline{n}\underline{n}:\underline{\underline{e}}_p)^2. \quad (36_{1,2})$$

Both these functions are positively semi-definite and do not exceed their corresponding values for isotropic case with the same modulus and viscosity.

In order to use in the completely soft case the evolution equation ($29_2$) for elastic strain $\underline{\underline{\varepsilon}}$, one should resolve uncertainties in expressions for $s_1$ and $s_2$ defined in (26). Employing for this purpose the formulae ($34_{1,2}$) for semi-soft cases in the limit $\{\delta_G, \delta_\eta, \kappa_G, \kappa_\eta\} \to 0$, one can formally reduce ($29_2$) to:

$$\begin{gathered}\overset{0}{\underline{\underline{\varepsilon}}} + s_0\mathbf{A}(\underline{n}; 1, r_1 - 1, 3(r_2 - 1)/2)\bullet\underline{\underline{\varepsilon}} = \underline{\underline{e}} \\ r_1 = \lim_{\delta_G, \delta_\eta \to 0}(\delta_G/\delta_\eta) > 0, \quad r_2 = \lim_{\kappa_G, \kappa_\eta \to 0}(\kappa_G/\kappa_\eta) > 0, \quad s_0 = G_0/\eta_0\end{gathered} \qquad (37_1)$$

Equation ($37_1$) described the *regular case* when two finite numerical values $r_1$ and $r_2$: ($0 < r_1, r_2 < \infty$) exist. In this case there is also the equivalent presentation of the evolution equation for the transient strain $\underline{\underline{\varepsilon}}$ [Leonov (2004)]:

$$\theta_0\mathbf{A}(\underline{n}; 1, r_1^{-1} - 1, 3(r_2^{-1} - 1)/2)\bullet(\underline{\underline{e}} - \overset{0}{\underline{\underline{\varepsilon}}}) = \underline{\underline{\varepsilon}} \qquad (37_2)$$

Using the equivalent evolution equations ($37_{1,2}$) it is now possible to consider the limiting cases when one of parameters $r_1$ or $r_2$ (or both) tends to zero or infinity. If $r_1$ (or $r_2$, or both) goes to zero, only the presentation ($37_1$) is valid. On the contrary, if $r_1$ (or $r_2$, or both) goes to infinity, only the presentation ($37_2$) is valid. Note that the simultaneous limits $r_1 \to 0$, $r_2 \to \infty$, or *vice versa*, $r_1 \to \infty$, $r_2 \to 0$, do not exist.

Consider now the possible *super-soft* cases, when both $r_1$ and $r_2$ simultaneously go either to zero (case1) or to infinity (case 2). Using the numerical operator $\boldsymbol{\alpha}(\underline{n})$ defined in ($35_3$), the evolution equations for elastic strain $\underline{\underline{\varepsilon}}$ in these two super-soft cases are:

$$\overset{0}{\underline{\underline{\varepsilon}}} + s_0\boldsymbol{\alpha}(\underline{n})\bullet\underline{\underline{\varepsilon}} = \underline{\underline{e}}, \qquad \boldsymbol{\alpha}(\underline{n})\bullet(\overset{0}{\underline{\underline{\varepsilon}}} - \underline{\underline{e}}) + s_0\underline{\underline{\varepsilon}} = 0. \qquad (38_{1,2})$$

Consider finally the evolution equation (30) for director in the complete soft cases. Noting that the only parameter $r_1$ is involved there in the regular case, the parameter $\theta^*$ in (30) is given by: $\theta^* = \eta_0/(G_0 r_1)$. Therefore in the two super-soft cases, the evolution equation (30) for director is reduced to the Ericksen equation, $\overset{0}{\underline{n}} = \lambda\mathbf{b}(\underline{n})\bullet\underline{\underline{e}}$, where $\lambda = \lambda_e$ or $\lambda = \lambda_v$ for the super-soft cases 1 or 2, respectively.

It should be noted that using the approximation of differentiation with frozen value of director, valid also for soft cases, one can also present the CE's ($37_{1,2}$) and ($38_{1,2}$) as the stress evolution equations with substitutions $\underline{\underline{\varepsilon}} \to \underline{\underline{\sigma}}$, $\underline{\underline{e}} \to \mathbf{G}(\underline{n})\bullet\underline{\underline{e}}$.

### 3.3. Linear symmetric nematic viscoelasticity

The CE's for symmetric nematic viscoelasticity derived in the previous Sub-Section, are weakly nonlinear. The linear case for equations (29) is achieved near the rest state with known value of $\underline{n}$, when a small gradient velocity is applied. This case is described by the linear equations (29) or by a single linear equation (31) where $\overset{0}{\underline{\underline{\sigma}}} \to \dot{\underline{\underline{\sigma}}}$, whereas the equation (30) for evolution of director is reduced to a linear equation for the small director disturbance $\delta \underline{n}$. Note that the pre-orientation in this case can also be made by a preliminary flow (say, pre-shearing). Thus the linear CE's can be introduced only when $\underline{n} = const$. In this case the exact solutions exist in both non-symmetric and symmetric *mono-domain* cases [Leonov (2004)]. We demonstrate below only a simple solution for symmetric case. When $\underline{n} = const$ the closed set of CE's is of the form ($29_{1,2}$).

A good example of linear behavior is the *Miesowicz' viscoelasticity*. It occurs when a relatively strong magnetic field $\underline{H}$ causes much higher effect on director orientation than the effects viscoelastic forces. In this case the rigid segments of macromolecules in LCP's are almost oriented along the magnetic field. Then due to (6), (7) and (19), one can establish the (Miesowicz') approximate relations:

$$\underline{n} = \underline{H}/|\underline{H}| + \delta\underline{n}, \quad \underline{h}_+ = (\chi_a H^2/G_5)\underline{n} + \underline{O}(\delta\underline{n}), \; \underline{h}_+^{\perp} = \underline{O}(\delta\underline{n}), \; ((\underline{\underline{h}}_+^a, \underline{\underline{h}}_+^s) = \underline{\underline{O}}(\delta\underline{n}). \quad (39_{1,2})$$

Here $|\delta\underline{n}| << 1$. Using (39) in evolution equation (22) one can see that the Miescowicz approximation holds for very slow flows when $(\underline{\underline{e}}, \underline{\underline{\omega}}, \underline{\underline{\varepsilon}}) = \underline{\underline{O}}(\delta\underline{n})$. Neglecting the small term $\delta\underline{n}$ in equations (21), (24), and (29) yields the CE's for Miescowicz viscoelasticity:

$$\dot{\underline{\underline{\varepsilon}}} + \mathbf{s}(\underline{n}) \bullet \underline{\underline{\varepsilon}} = \underline{\underline{e}}, \quad \underline{\underline{\sigma}} = \mathbf{G}(\underline{n}) \bullet \underline{\underline{\varepsilon}} \qquad (40_{1,2})$$

In the evolution equation ($40_1$) for elastic strain $\underline{\underline{\varepsilon}}$, the overdot means the (partial) time derivative. Considering below only the space homogeneous magnetic field results in the condition: $\underline{n} = \underline{H}/|\underline{H}| = const$. The set ($40_{1,2}$) presents the linear and symmetric, nematic viscoelastic CE's. Another possibility when the linear equations (40) are valid is pre-shearing.

Consider now the problem of solving equations ($40_{1,2}$) when the strain rate $\underline{\underline{e}}$ is given. A new effective algebraic technique for solving this problem has been elaborated [Leonov (2004)] that presents the solution in terms of linear memory functionals of $\underline{\underline{e}}$. This approach is briefly described as follows. Searching the solution of homogeneous linear ODE ($30_1$) in the form, $\underline{\underline{\varepsilon}} = \underline{\underline{\psi}} \exp(-\nu t)$, reduces it to the common spectral problem, $\mathbf{s}(\underline{n}) : \underline{\underline{\psi}} - \nu \underline{\underline{\psi}} = \underline{\underline{0}}$, where $\nu$ and $\underline{\underline{\psi}}$ are the eigenvalue and corresponding "eigentensor". The eigenvalues $\nu_k$ of the spectral problem are found as:

$$\nu_1 = s_0 = \frac{G_0}{\eta_0}, \quad \nu_2 = s_0 + s_1 = \frac{G_0 + G_1}{\eta_0 + \eta_1}, \; \nu_3 = s_0 + \frac{2}{3}s_2 = \frac{3/4G_0 + G_1 + G_2}{3/4\eta_0 + \eta_1 + \eta_2}. \qquad (41)$$

Due to the thermodynamic stability conditions ($27_{1,2}$) all the eigenvalues $\nu_k$ are positive and describe the *relaxation frequencies*, with the respective *relaxation times* $1/\nu_k$. The new algebraic technique also allows the integral presentation of the problem ($31_1$) in a very brief way:

$$\underline{\underline{\varepsilon}} = \mathbf{A}(\underline{n}; \underline{\underline{X}}_1, \underline{\underline{X}}_2, \underline{\underline{X}}_3), \qquad \underline{\underline{X}}_k(t) = \chi_k(t) * \underline{\underline{e}}(t)$$
$$\chi_0(t) = e^{-\nu_1 t}, \; \chi_1(t) = e^{-\nu_2 t} - e^{-\nu_1 t}, \; \chi_2(t) = \frac{3}{2}(e^{-\nu_3 t} - e^{-\nu_1 t}) \qquad (42)$$

Hereafter the convolution of two functions in interval $(-\infty, t)$ are defined as:

$$\{f(t)*\varphi(t)\}\equiv\int_{-\infty}^{t} f(t-\tau)\varphi(\tau)d\tau$$

Substituting (42) into ($29_2$) yields the stress-strain rate relation presented in the form of memory functionals:

$$\underline{\underline{\sigma}}(t)=\underline{\underline{E}}_0(t)+\underline{nn}\cdot\underline{\underline{E}}_1(t)+\underline{\underline{E}}_1(t)\cdot\underline{nn}-2\underline{nn}[\underline{\underline{E}}_1(t):\underline{nn}]+[\underline{\underline{E}}_2(t):\underline{nn}](\underline{nn}-1/3\underline{\underline{\delta}})$$
$$\underline{\underline{E}}_k(t)\equiv\{m_k(t)*\underline{\underline{e}}(t)\} \quad . \qquad (43)$$

Here the kernels $m_k(t)$ $(k=0,1,2)$ in the memory functionals $\underline{\underline{E}}_k(t)$ in (43) are:

$$m_0(t)=G_0e^{-\nu_1 t},\ \ m_1(t)=-G_0e^{-\nu_1 t}+(G_0+G_1)e^{-\nu_2 t}$$
$$m_2(t)=-3/2G_0e^{-\nu_1 t}+2(3/4G_0+G_1+G_2)e^{-\nu_3 t} \qquad (44)$$

Equations (43), (44) derived here from the linear differential CE's ($29_{1,2}$) are of the type, proposed *ad hock* by Larson and Mead (1989).

There are two physically evident limits of the memory functionals $\underline{\underline{E}}_k(t)$.

1) When $t<<\min\nu_k^{-1}=\min\hat{\theta}_k$, one has: $\underline{\underline{E}}_k(t)\approx m_k(0)\int_{-\infty}^{t}\underline{\underline{e}}(\tau)d\tau=G_k\underline{\underline{\varepsilon}}(t)$. Here $G_k$ are the basis moduli in ($31_2$). This is the elastic limit of (42) which coincides with ($29_2$).

2) When $t\to\infty$ and the constant limit $\underline{\underline{e}}\equiv\underline{\underline{e}}(\infty)$ exists, $\underline{\underline{E}}_k(\infty)=\underline{\underline{e}}\int_0^{\infty}m_k(t)dt=\eta_k\underline{\underline{e}}$, where $\eta_k$ are the basic viscosities. In this case the relations (42), (43) are purely viscous and present the limit: $\underline{\underline{e}}_p\to\underline{\underline{e}}$.

In the completely soft cases when all the equalities (34) are satisfied, and generally, $\nu_1=s_0=G_0/\eta_0$, $\nu_2=r_1s_0$, $\nu_3=r_2s_0$ $(0<r_1,r_2<\infty)$, the functions $\chi_k(t)$ in (42) become:

$$\chi_0(t)=e^{-s_0t},\ \chi_1(t)=e^{-r_1s_0t}-e^{-s_0t},\ \chi_2(t)=\frac{3}{2}(e^{-r_2s_0t}-e^{-s_0t})\,.$$

Here the super-soft cases 1 and 2 are respectively achieved when $r_1,r_2\to0$ or $r_1,r_2\to\infty$.

In the completely soft case independently of parameters $r_1$ and $r_2$, $m_0(t)=G_0e^{-s_0t}$, $m_1(t)=-G_0e^{-s_0t}$ and $m_2(t)=-3/2G_0e^{-s_0t}$. Then the memory functionals (43) are:

$$\underline{\underline{\sigma}}/G_0=\boldsymbol{\alpha}(\underline{n})\bullet\underline{\underline{E}}(t),\quad \underline{\underline{E}}(t)\equiv\int_{-\infty}^{t}\exp[-s_0(t-\tau)]\underline{\underline{e}}(\tau)d\tau\,, \qquad (45)$$

Here $s_0=G_0/\eta_0$ is the relaxation frequency (reciprocal relaxation time), and the right-hand side of the first equality in (45) is defined in ($37_3$). Note that Eq. (45) predicts the existence of both the completely viscous soft and viscoelastic super-soft deformation modes when the only conditions are satisfied in (33) for existence of complete elastic soft mode, $G_0+G_1=0$ and $3/4G_0+G_1+G_2=0$.

## IV. Examples: simple shearing and simple elongation

This Section illustrates some flow effects of nematics in simple shearing and simple elongation, described by several analytical solutions of weakly nonlinear viscoelastic nematodynamic CE's. The simplified set of CE's consisting of the evolution equations for director (30) and for extra stress ($31_2$) is used in this Section.

4.1. Simple shearing.

The shearing flows is analyzed utilizing the common Cartesian coordinate system $\{\underline{x}\} = \{x_1, x_2, x_3\}$ where $x_1$ is directed along the flow and $x_2$ along the velocity gradient. In this coordinate system, the tensors of strain rate $\underline{\underline{e}}(t)$ and vorticity $\underline{\underline{\omega}}(t)$ for homogeneous shearing flows have the matrix forms:

$$\underline{\underline{e}}(t) = \dot{\gamma}(t)\underline{\underline{\alpha}}, \quad \underline{\underline{\omega}}(t) = \dot{\gamma}(t)\underline{\underline{\beta}}, \quad \underline{\underline{\alpha}} = \frac{1}{2}\begin{pmatrix} 0 & 1 & 0 \\ 1 & 0 & 0 \\ 0 & 0 & 0 \end{pmatrix}, \quad \underline{\underline{\beta}} = \frac{1}{2}\begin{pmatrix} 0 & -1 & 0 \\ 1 & 0 & 0 \\ 0 & 0 & 0 \end{pmatrix}. \tag{46}$$

*4.1.1. Linear dynamic properties*

Here $\dot{\gamma}(t) = \gamma_0 e^{i\omega t}$, with $\gamma_0$ being the small amplitude and $\omega$ the frequency of shearing oscillations. Substituting (46) into (43) with account for (44) yields:

$$\underline{\underline{\mathbf{G}}}^*(\omega) = \mathbf{A}(\underline{n}; G_1^*, G_2^*, G_3^*) : \underline{\underline{\alpha}}. \tag{47}$$

Here $\underline{\underline{\mathbf{G}}}^*(\omega) = \underline{\underline{\sigma}}/\gamma$ is the dynamic complex tensor-modulus, and $\mathrm{G}_k^*(\omega)$ are the basic scalar complex moduli:

$$G_1^*(\omega) = \frac{G_0 i\omega}{2(1+i\omega\nu_1)}, \quad G_2^*(\omega) = \frac{(G_0+G_1)i\omega}{2(1+i\omega\nu_2)}, \quad G_3^*(\omega) = \frac{2(3/4G_0+G_1+G_2)i\omega}{2(1+i\omega\nu_3)}. \tag{48}$$

Note that in the completely soft elastic case $G_2^*(\omega) = G_3^*(\omega) = 0$.

When the director $\underline{n}$ is disposed arbitrarily relative to the coordinate system $\{\underline{x}\}$, the components of complex tensor modulus $\underline{\underline{\mathbf{G}}}^*(\omega)$ are:

$$\begin{aligned}
&\mathbf{G}_{12}^*(\omega) = (1+n_1^2n_2^2-n_1^2-n_2^2)G_1^* + (n_1^2+n_2^2-4n_1^2n_2^2)G_2^* + 2n_1^2n_2^2G_3^* \\
&\mathbf{G}_{13}^*(\omega) = n_3[(n_1-n_2+n_1^2n_2)G_1^* + n_2(1-4n_1^2)G_2^* + 2n_1^2n_2G_3^*(\omega)] \\
&\mathbf{G}_{23}^*(\omega) = n_3[(n_2-n_1+n_1^2n_2)G_1^* + n_1(1-4n_1n_2)G_2^* + 2n_1n_2^2G_3^*] \\
&\mathbf{G}_{n1}^* \equiv \mathbf{G}_{11}^* - \mathbf{G}_{22}^* = (n_1^2-n_2^2)[(1+n_1n_2)G_1^* - 4n_1n_2G_2^* + 2n_1n_2G_3^*] \\
&\mathbf{G}_{n2}^* \equiv \mathbf{G}_{22}^* - \mathbf{G}_{33}^* = [(n_2^2-n_3^2)(1+n_1n_2) - 2n_1n_2]G_1^* - 2n_1n_2[2(n_2^2-n_3^2)-1]G_2^* \\
&\qquad + 2n_1n_2(n_2^2-n_3^2)G_3^* .
\end{aligned} \tag{49}$$

In the particular case with the director located in the $\{x_1, x_2\}$ shearing plane when $n_1 = \cos\theta$, $n_2 = \sin\theta$, $n_3 = 0$, non-zero components of the tensor complex modulus are:

$$\begin{aligned}
&\mathbf{G}_{12}^*(\omega) = 1/4G_1^*\sin^2 2\theta + G_2^*\cos^2 2\theta + 1/2G_3^*\sin^2 2\theta \\
&\mathbf{G}_{n1}^*(\omega) = G_1^*\cos 2\theta + (G_1^*/4 - G_2^* + G_3^*/2)\sin 4\theta \\
&\mathbf{G}_{n2}^*(\omega) = G_1^*[(\sin^2\theta - \sin 2\theta(1-1/2\sin^2\theta)] - G_2^*\sin 2\theta(2-\sin^2\theta) + G_3^*\sin^2\theta\cdot\sin 2\theta
\end{aligned} \tag{50}$$

Here the particular cases $\theta = 0$ and $\theta = \pi / 2$ of the pre-oriented nematics in directions of either flow or velocity gradient could be practically realized by imposing magnetic field. The general case $0 < \theta < \pi / 2$ in (50) seems to be more realistic for description of pre-sheared dynamic experiments of aligning nematics (see the Subsection 4.1.2. below). The possible cases of softness/semi-softness could also play important role in experimental verification of these predictions. It should also be mentioned that for the case $n_3 = 1,\ n_1 = n_2 = 0$ when the director is orthogonal to the flow plane $\{x_1, x_2\}$, the only non-zero component of the complex tensor modulus is $\mathbf{G}^*_{n2}(\omega) = -G^*_1(\omega)$.

We now illustrate some *weakly nonlinear effects* in simple shearing.

*4.1.2. Evolution of director located in the $\{x_1, x_2\}$ shearing plane*

Substituting expression for $\underline{n} = \{n_1, n_2, 0\}$ and kinematical matrices (46) into evolution equation for director (30) yields the evolution equation for the longitudinal component $n_1$ of director:

$$\theta^*(\ddot{n}_1 + n_1\dot{n}_1^2 / n_2^2) + \dot{n}_1 - \dot{\gamma} n_2 / 2 = \lambda_e \theta^* \ddot{\gamma}(1 - 2n_1^2) n_2 / 2 - \lambda_e \theta^* \dot{\gamma}^2 n_1 n_2^2 + \lambda_v \dot{\gamma} n_2 (1 - 2n_1^2) / 2 \quad (51)$$

Here $n_2 = \sqrt{1 - n_1^2}$.

We now analyze the possible case of *aligning nematics* in steady shearing, when $\dot{\gamma} = const$. Assuming that $n_1 = const$ equation (51) yields:

$$\sqrt{1 - n_1^2} = 0;\ \ 1 - 2\lambda_e D n_1 \sqrt{1 - n_1^2} + \lambda_v (1 - 2n_1^2) = 0 \quad (D = \theta^* \dot{\gamma}). \quad (52)$$

Hereafter $D$ is the flow Deborah number. Along with trivial solution $n_1 = \pm 1$, showing the completely aligning case, there is the non-trivial $D$-dependent solution of (52):

$$n_1^2 = \frac{1}{2}\left(1 + \lambda_v \frac{1 + |D\lambda_1 / \lambda_v| \sqrt{D^2 \lambda_1^2 + \lambda_v^2 - 1}}{D^2 \lambda_1^2 + \lambda_v^2}\right). \quad (53)$$

Formula (53) covers both the limiting solutions of the second equation in (52), $n_1 = 0$ and $n_1 = \pm 1$, when $\lambda_v \to \pm 1$, respectfully. The aligning case exists if $0 \le n_1^2 \le 1$. The criterion of existence is readily established as:

$$r^2 \equiv D^2 \lambda_e^2 + \lambda_v^2 \ge 1. \quad (54)$$

If (54) is satisfied, introducing in (53) the new variables, $r^{-1} = \cos\phi$, $D\lambda_e = r\cos\psi$, $\lambda_v = r\sin\psi$, the short calculations below show that the second term in bracket in (53) does not exceed the unity:

$$\lambda_v \frac{1 + |D\lambda_e / \lambda_v| \sqrt{D^2 \lambda_e^2 + \lambda_v^2 - 1}}{D^2 \lambda_e^2 + \lambda_v^2} = r^{-1}\sin\psi \pm \sqrt{1 - r^{-2}}\cos\psi = \cos\phi \sin\psi \pm \sin\phi \cos\psi = \sin(\psi \pm \phi).$$

Here $\pm$ correspond to the $\pm$ signs of $\lambda_v$. Formula (53) shows that in very slow flow when $D << 1$, the criterion of aligning case to exist, $|\lambda_v| > 1$, coincides with that known for viscous (low molecular weight) nematic fluids. Nevertheless, at higher Deborah number, the model predicts the existing of aligning flow situation even if $|\lambda_v| < 1$. Although the non-aligning case, $r^2 < 1$, is not analyzed in the present paper, one may assume that it describes the *tumbling* effects with self-oscillations of director. Therefore one may speculate that the viscoelastic effects in inherently non-aligning case $|\lambda_v|<1$ may suppress the oscillations and stabilize the shearing flow. Formulae (53) demonstrate that shear dependent orientation of director, proportional to $|\lambda_e|$, occurs only due to effect of *elastic internal rotations* (see (34)).

Equation (53) can be used for making expansions for *D*-dependent orientation in low Deborah number flow, $D<<1$. These expansions, valid with the accuracy of $O(D^2)$, are:

$$n_1^2 \approx \frac{1}{2}\left(1+\lambda_v^{-1}+\lambda_v\left|D\lambda_e/\lambda_v\right|\sqrt{1-\lambda_v^{-2}}\right),\ \ n_2^2 \approx \frac{1}{2}\left(1-\lambda_v^{-1}-\lambda_v\left|D\lambda_e/\lambda_v\right|\sqrt{1-\lambda_v^{-2}}\right). \qquad (55_1)$$

Although the present weakly nonlinear theory cannot describe the high Deborah number flows, we also present the formal expansions of (53) for $D\to\infty$, valid with accuracy of $O(D^{-2})$:

$$n_1^2 \approx \frac{1+\operatorname{sgn}\lambda_v}{2}-(\operatorname{sgn}\lambda_v)\frac{(|\lambda_v|-1)^2}{4D^2\lambda_v^2},\ \ n_2^2 \approx \frac{1-\operatorname{sgn}\lambda_v}{2}+(\operatorname{sgn}\lambda_v)\frac{(|\lambda_v|-1)^2}{4D^2\lambda_e^2}. \qquad (55_2)$$

When $\lambda_v>0$ the formulae ($55_2$) describe the hypothetical case when at $D\to\infty$ the director aligns along the velocity direction. When $\lambda_v<0$, formulae ($55_2$) describe the case when the director aligns along the gradient velocity direction. Although the second option looks unstable, our CE's are invalid in this limit.

Consider now another example, the evolution of director after cessation of steady shearing $(\dot\gamma=0)$ with aligning director. Introducing in (51) $n_1=\cos\varphi,\ n_2=\sin\varphi$, reduces (51) to the relaxation equation, $\theta^*\ddot\varphi+\dot\varphi=0$, whose solution is:

$$\varphi(t)=\varphi_0+D\lambda_e(1-e^{-t/\theta^*})\cos 2\varphi_0. \qquad (56)$$

Here the value $\varphi_0(D)$ is known from the steady shearing, the initial value $\dot\varphi_0=\dot\gamma\lambda_e\cos 2\varphi_0$ being readily established from (51) using the jump condition at the instant of director relaxation.

*4.1.3. Steady shearing of aligning viscoelastic nematics*

After tedious calculations, the steady CE's ($31_2$) result in the linear set of algebraic relations for the shear stress $\sigma_{12}$ and the first $N_1$ and second $N_2$ normal stress differences:

$$\begin{aligned}
&N_1\dot\gamma(\theta^*/2+\theta_\beta n_1^2n_2^2)+\sigma_{12}[1+\theta_\beta\dot\gamma n_1n_2(1-2n_1^2)]=\dot\gamma(\eta_s+2\eta_2n_1^2n_2^2)\\
&N_1[1-\theta_\beta\dot\gamma n_1n_2(1-2n_1^2)]-\dot\gamma\sigma_{12}(\theta_\beta+2\theta^*-4\theta_\beta n_1^2n_2^2)=-2\dot\gamma\eta_2n_1n_2(1-2n_1^2) \qquad (57)\\
&N_2=\dot\gamma n_1n_2[\eta_1+2\eta_2n_2^2+N_1(\theta_2-\theta_1-\theta_\beta n_1^2)]-\dot\gamma\sigma_{12}(\theta_0+\theta_2n_2^2-2\theta_\beta n_1^2n_2^2)
\end{aligned}$$

Here $n_1$ and $n_2$ depend on the flow rate as shown in (53), and

$$\theta^*=\theta_0+\theta_1,\quad \theta_\beta=\theta_2-2\theta_1,\quad \eta_s=(\eta_0+\eta_1)/2. \qquad (58)$$

The general solution of first two equations is rather awkward. Nevertheless, in the *complete soft* (*viscous and* elastic) case, when $\eta_s=0$ and $\eta_2=\eta_0/4$, this solution is:

$$\sigma_{12}=\frac{\eta_0\dot{\gamma}n_1n_2[2n_1n_2-D(n_1^2-n_2^2)]}{4(1+D^2q)},\quad N_1=\frac{\eta_0\dot{\gamma}n_1n_2(n_1^2-n_2^2+2Dn_1n_2)}{2(1+D^2q)},\quad q=r_1\frac{r_2+3}{4r_2}. \qquad (59)$$

Here $D=\theta^*\dot{\gamma}$, $\theta^*=\theta_0/r_1$, and the positive parameters $r_1$ and $r_2$ have been introduced in the complete soft case in equations ($37_{1,2}$). The expression for $N_2$ in this case, though more awkward, is also easy to obtain. In the "super-soft" limit $r_1\to\infty$, when $\theta^*\to 0$, formulae (59) are reduced to those described the soft nematic viscous case with constant values of director orientation depending only on viscous tumbling parameter $\lambda_v$ [see Leonov and Volkov (2005)].

Using expansions ($55_1$), the weakly non-Newtonian case of (59) is presented with accuracy of terms $O(D^2)$ as:

$$\sigma_{12}\approx\eta_N\dot{\gamma}\left(1-2\frac{|\lambda_e D/\lambda_v|}{\sqrt{1-\lambda_v^{-2}}}\right),\quad N_1\approx\frac{4\eta_N\dot{\gamma}}{\lambda_v\sqrt{1-\lambda_v^{-2}}}\left(1+|D\lambda_e/\lambda_v|\frac{1+|\lambda_v|(\lambda_v^2-1)}{\sqrt{\lambda_v^2-1}}\right). \qquad (60)$$

Here $\eta_N=\eta_0(1-\lambda_v^{-2})/8$ is the Newtonian shear viscosity. The first formula in (60) demonstrates that the apparent shear viscosity is shear thinning. The sign of the first normal stress difference $N_1$ depends on the sign of the viscous tumbling parameter $\lambda_v$, $|N_1|$ being increasing with $D$ increasing. It also seen from (59) that at high values of Deborah number, the model collapses without physical reasons, exactly as in the isotropic viscoelastic case.

4.2. Simple elongation

In simple elongation, $\underline{\underline{\omega}}=0$, and in the common orthogonal coordinate system $\{\underline{x}\}=\{x_1,x_2,x_3\}$ where $x_1$ is directed along the flow, $x_2$ is normal to flow direction and located in the meridian plane, the strain rate $\underline{\underline{e}}(t)$ is presented in the matrix form as:

$$\underline{\underline{e}}(t)=\dot{\varepsilon}(t)\underline{\underline{\kappa}},\quad \underline{\underline{\kappa}}=\begin{pmatrix}1&0&0\\0&-1/2&0\\0&0&-1/2\end{pmatrix}. \qquad (61)$$

The evolution equation for director (30) takes the form:

$$\theta^*(\underline{\ddot{n}}+\underline{n}|\underline{\dot{n}}|^2)+\underline{\dot{n}}=\frac{3}{2}(\lambda_e\theta^*\ddot{\varepsilon}+\lambda_v\dot{\varepsilon})(\underline{v}-n_1^2\underline{n}),\quad \underline{v}=(n_1,0,0). \qquad (62)$$

In the following, the evolution equation is analyzed in case when director is located in the elongation plane $\{x_1,x_2\}$. Substituting into (62) the kinematical matrices (61) and $\underline{n}=\{n_1,n_2,0\}$ where $n_1=\cos\varphi,\ n_2=\sin\varphi$, yields the evolution equation for director orientation:

$$\sin 2\varphi[\theta^*\ddot{\varphi}+\dot{\varphi}+(\theta^*\dot{u}+u)\sin 2\varphi]=0,\quad u=(3/4)\lambda_v\dot{\varepsilon}. \qquad (63)$$

Consider first the evolution of 2D director in start up elongation flow with $\dot{\varepsilon}=const>0$, when $\dot{u}=0$. Along with trivial solutions of (63), $\varphi=0,\ \varphi=\pi/2\ (\pm\pi n)$, there also exists the non-trivial solution of equation in the bracket of (63). Since in the simple extension $\dot{\varepsilon}>0$, the sign of parameter $u$ is pre-determined by the sign of the viscous tumbling parameter $\lambda_v$. The analysis of linear disturbances near $\varphi = 0$ ($\pm \pi n$) aligned in flow direction, shows that the point $\varphi = 0$ is attractor if $u > 0$ (or $\lambda_v > 0$). In this case the disturbances aperiodically decay when $4u\theta^* < 1$, and decay periodically otherwise. When $u < 0$ (or $\lambda_v < 0$), the substitution $\varphi = \pi/2 - \varphi_1$ reduces the linear analysis of disturbances to the previous one. It means that when $\lambda_v < 0$, the 2D orientation of director (located in the meridian plane) normal to the flow direction is 2D stable. Phase diagram for the solution of nonlinear problem (63) confirms these results of linear disturbance analysis. It should be noted that these results qualitatively coincides with the prediction of nonlinear 2D solution of Ericksen orientation problem ($\theta^* = 0$):

$$\frac{n_1^2}{1-n_1^2} = \frac{(n_1^0)^2 e^{ut}}{1-(n_1^0)^2}. \tag{64}$$

Here $n_1(t) = \cos\varphi(t)$ and $n_1^0 = \cos\varphi(0)$ are the actual and initial 2D orientation of director. Formula (64) shows that $n_1 \to \pm 1$ when $u > 0$, and $n_1 \to 0$ when $u < 0$. It seems that the flow with director aligning in direction normal to flow is 3D unstable.

As in the simple shearing, the evolution of director after cessation of an elongation flow is again described by the equation, $\theta^* \ddot{\varphi} + \dot{\varphi} = 0$, whose solution is:

$$\varphi(t) = \varphi_0 - u_0(1 - e^{-t/\theta^*})\sin 2\varphi_0. \tag{65}$$

Here $\varphi_0$ and $u_0$ are determined at the instant $t-0$ prior the director relaxation. The relation (65) shows that in the limit cases, $\varphi_0 = 0$ and $\varphi_0 = \pm\pi/2$, the possible director relaxation might be only 3-dimensional.

We finally consider the steady elongation flow. According to ($31_2$) in this case, the stress-strain rate relation predicted by the model is viscous, $\underline{\underline{\sigma}} = \boldsymbol{\eta}(\underline{n}) \bullet \underline{\underline{e}}$, with $n_1$ being equal either to $\pm 1$ or to $0$ for director respectfully aligning either along or normal to the flow direction. The elongation viscosity for both the cases is given by

$$\eta_{el} = \sigma_{el}/\dot{\varepsilon} = 3\eta_0/2 + 6\eta_1 n_1^2 n_2^2 + (\eta_1 + \eta_2)(n_1^2 - n_2^2)(2n_1^2 - n_2^2). \tag{66}$$

The relation (66) describes both the types of director aligning:

$$\eta_{el}\Big|_{n_1=1} = 3\eta_0/2 + 2(\eta_1 + \eta_2), \quad \eta_{el}\Big|_{n_1=0} = 3\eta_0/2 + \eta_1 + \eta_2. \tag{67}$$

In the complete soft case, when $\eta_1 = -\eta_0$, $\eta_1 + \eta_2 = -3\eta_0/4$, the formulae in (67) yield: $\eta_{el} = 0$ if the director is oriented along the flow ($n_1 = \pm 1$) [Leonov and Volkov (2005)], or $\eta_{el} = 3\eta_0/4$ if the director is oriented perpendicular to flow ($n_1 = 0$). Note that the dissipation is less in the first case than in the second one.

.

## V. Conclusions

**1.** This paper developed a continual theory of weak viscoelastic nematodynamics in the simplest case of Maxwell nematic fluid model. It can describe the molecular elasticity effects in mono-domain flows of LCP's as well as the viscoelastic effects in slow flows of suspensions of uniaxially symmetric particles in polymer fluids. The proposed phenomenology started with well-known fractioning of total deformations/rotations in the transient (elastic) and irreversible (viscous) parts, which is indifferent to the particular rheology of continuum. Additional kinematical relations that describe the internal rotations are specific for nematic continua. Weakly nonlinear character of this theory is based on the assumption of smallness of elastic (transient) strains and elastic relative rotations. It means in fact the smallness of the Deborah numbers, with valid in this case the co-rotational tensor time derivatives. The theory utilized the CE's resulting from the de Gennes-type potential for weakly elastic nematic solids in the presence of magnetic field, and the LEP-type CE's for viscous nematic liquids, while ignoring the Frank (orientation) elasticity and inertia effects. It seems that the director field in this theory cannot be included in the set of state variables, as was recently discussed for viscous nematic liquids [Leonov (2005)]. This is because the evolution of director is described by the external fields and, due to additional nematic kinematics, by flow field. The stress asymmetry is demonstrated on the example of action of magnetic field. To obtain the evolution equations for tensor state variables, such as the transient strain and transient relative rotation, one need to exclude from the initial formulation the non-equilibrium rates of deformation and relative rotations. This operation, trivial for isotropic viscoelasticity, is complicated in general multi-parametric case of nematic viscoelasticity. Therefore a new general algebraic approach to nematic operations was recently developed [Leonov (2004)], which revealed a general structure of the theory. The coupled evolution equations obtained in such a way for the tensor state variables, resulted also in the viscoelastic character of evolution equation for director.

**2.** In the absence of magnetic field, when the stress tensor is symmetric and orientation of director caused only by flow, the theory has the transversally anisotropic viscoelastic characte. In this case, the simplified approach resulted in a closed set of two coupled anisotropic visoelastic equations (30) and (31) for evolution of director and extra stress.

The possible soft nematic deformation modes have been completely analyzed in the weak elastic and viscous cases by Leonov and Volkov (2004a,b), using the marginal stability approach. It was shown that the soft modes could exist only in the absence of external fields. In these papers, the deep analogy between the behavior of magnetics and nematic elastic solids, established in the Golubovich-Lubensky (1989) conjecture, was also extended to the viscous case [see also Leonov (2005)]. The general physical principle assumed in these papers is that in the nematic systems, large fluctuations could bring the system to the marginal state with absolute minimum of free energy [Lubensky and Mukhopadya (2002)] or dissipation. The energy barrier needed for this minimum to be located close to the marginal state can be provided by various small stabilizing terms, such as the Born term typically neglected in nematic theories. In the present paper, the general formal analysis of possible soft modes [Leonov (2004)] is demonstrated for a particular case when both the viscous and elastic modes in the model are soft in shearing and elongation deformations. As usual, the soft modes highly reduce the number of constitutive parameters, and in some cases, demonstrate the "nematic superfluidity" effects where in ideal no energy is needed to maintain the flow.

The linear nematic symmetric viscoelasticity was then investigated and exemplified on the example of "Miescowicz viscoelasticity". Using the algebraic approach [Leonov (2004)] the linear CE's were presented via linear memory fucntionals depending on the strain rate.

**3.** Some analytical predictions of the theory were demonstrated using the simplified CE's (30) and (31) for simple shearing and simple elongation flows in the absence of magnetic field.

For *simple shearing*, the general dynamic low amplitude formulae were presented when the preliminary established value of director is arbitrarily disposed relative the shearing plane, and then simplified for the case when the 2D director is located in the shearing plane. These formulae will further be applied for comparisons of the predictions with experiments. The theory also predicts the shear dependence of 2D director in aligning case, which is dominant in the weak non-Newtonian behavior of the viscoelatic nematics, with the sign of the first normal stress difference being equal to the sign of viscous tumbling parameter $\lambda_v$. Although the possible tumbling case has not been analyzed, the analysis predicts that the viscoelastic effects stabilize the director orientation even if the value of $\lambda_v$ belongs to unstable (non-align) region ($|\lambda_v| < 1$). We also demonstrated the relaxation effects of director evolution after cessation of steady shearing. Our approach predicts that the director relaxes its orientation gained in the previous shearing during the characteristic time $\theta^*$, which is of the same order of the stress relaxation time.

For *simple elongation*, the theory predicts that in the start up elongation flow, the director, being initially arbitrarily oriented, stabilizes in time either in the flow direction when $\lambda_v > 0$ or perpendicular to flow when $\lambda_v < 0$. These predictions qualitatively coincide with the solution of Ericksen orientation equation for viscous nematics. The expression for steady elongation viscosity shows that the first case with the soft mode demonstrates the elongational "superfluidity" effect with zero elongation viscosity, and second case the positive viscosity; both the values of elongation viscosity being less than the viscosity $\eta_0$ of isotropic term in nematic viscosity.

Detailed comparisons of this theory with experimental data need numerical calculations, which could be done after evaluation of material parameters.

## Acknowledgement

The author is thankful to Professor V.S. Volkov for valuable discussions.